\documentclass[twoside]{article}
\usepackage{fleqn,espcrc2}
\usepackage{graphicx}
\usepackage{amsmath}  
\usepackage{epsfig}

\newcommand{\newc}{\newcommand}
\newc{\alphastar}{\alpha_s(q^*)}
\newc{\alphastarn}[1]{{\alpha_s^{#1}}(q^*)}
\newc{\alphas}{\alpha_s}
\newc{\alphaq}{\alpha_s(q)}
\newc{\alphamu}{\alpha_s(\mu)}
\newc{\alphamutwo}{\alpha_s^2(\mu)}
\newc{\alphamuthree}{\alpha_s^3(\mu)}
\newc{\logqn}[1]{\log^{#1}(q^2)}
\newc{\logqm}{\log(q^2/\mu^2)}
\newc{\logqmtwo}{\log^2(q^2/\mu^2)}
\newc{\logqsm}{\log(q^{*2}/\mu^2)}
\newc{\bzero}{\beta_0}
\newc{\bone}{\beta_1}
\newc{\btwo}{\beta_2}
\newc{\bthree}{\beta_3}
\newc{\msbar}{{\scriptscriptstyle {\rm \overline{MS}}}}
\newc{\MSbar}{{\overline{\rm MS}}}
\newc{\dq}{d^4\!q\,}
\newc{\dnk}{d^n\!k\,}
\newc{\ave}{\int\dq}
\newc{\eq}[1]{Eq.~(\ref{#1})}
\newc{\Ref}[1]{Ref.~\cite{#1}}
\newc{\Refs}[1]{Refs.~\cite{#1}}
\newc{\fig}[1]{Fig.~\ref{#1}}
\newc{\Table}[1]{Table~\ref{#1}}
\newc{\sect}[1]{Sect.~\ref{#1}}
\newc{\vev}[1]{\left\langle {#1}\right\rangle}
\newc{\Vev}[1]{\big\langle\!\big\langle {#1}\big\rangle\!\big\rangle}
\newc{\Mb}{M_b}
\newc{\Mbmsbar}{\overline{M}_b}
\newc{\Mtau}{M_\tau}
\newc{\Mtaumsbar}{\overline{M}_\tau}
\newc{\Mt}{M_t}
\newc{\Mtmsbar}{\overline{M}_t}
\newc{\first}{1^{\rm st}}
\newc{\second}{2^{\rm nd}}

\title{Scale Setting for $\alpha_s$ Beyond Leading Order}

\author{K.~Hornbostel%
\address{Southern Methodist University, Dallas, TX 75275}%
, G.~P.~Lepage%
\address{Newman Laboratory of Nuclear Studies,
Cornell University, Ithaca, NY 14853}%
and 
C.~Morningstar%
\address{Carnegie Mellon University, Pittsburgh, PA 15213}%
\thanks{This work supported in part by the NSF and DOE.}
}

\begin{document}

\begin{abstract}
We present a general procedure for applying the scale-setting 
prescription of Brodsky, Lepage and Mackenzie to higher orders
in the strong coupling constant $\alphas$.  In particular, we
show how to apply this prescription when the leading coefficient
or coefficients in a series in $\alphas$ are anomalously small.  
We give a general method for computing an optimum scale numerically, 
within dimensional regularization, and in cases when the coefficients 
of a series are known.  
We find significant corrections to the scales for $R_{e^+ e^-}$,
$\Gamma(B \rightarrow X_u e \overline{\nu})$, 
$\Gamma(t \rightarrow b W)$, and the ratios of the quark pole to
$\MSbar$ and lattice bare masses.
\vspace{1pc}
\end{abstract}

\maketitle

\section{Introduction}

QCD processes computed to a finite order in perturbation theory
depend on the scale chosen for the running coupling constant $\alphaq$.  
While these variations diminish as higher orders are included, for low-order
calculations they can be significant.
Finding an optimum, physically motivated method 
for choosing this scale is important not only to produce
accurate results, but also to reasonably estimate convergence based 
on the size of the series coefficients.  Such a method allows a
meaningful prediction or comparison with data even at leading order.

In this paper, we apply the prescription defined by Brodsky, Lepage
and Mackenzie (BLM) in \Refs{Brodsky:1983gc,Lepage:1993xa} 
beyond leading order.  In so doing, we remedy an anomaly observed in a variety 
of applications as discussed, for example, in \Ref{Morningstar:1995vd}.  
We show that this requires a simple extension of the calculation needed 
to set the scale at lowest order.  We find that it leads to significant
corrections in the scales of several important processes, and allows us
to extract masses from lattice simulations which were previously
inaccessible.
(See \Refs{Beneke:1995qe,Neubert:1995vb} for other higher-order extensions.)

\section{The Prescription}

\begin{figure}[hbt]
\begin{center}
\epsfig{file=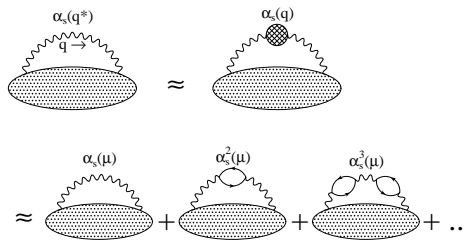, height=0.75in}
\end{center}
\caption{ 
  The BLM prescription fixing the optimum scale $q^*$ 
  to leading order in $\alphastar$.  Vacuum bubbles include both fermions and gluons.}
\label{fd1}
\end{figure}

The original statement of the prescription, 
\begin{equation}
\alphastar \ave f(q) \approx  \ave \alpha_s(q) f(q)
\end{equation}
illustrated in \fig{fd1}, is to choose the scale $q^*$ associated with a 
particular gluon such that it matches, as well as possible, the same diagram 
with a fully dressed gluon~\cite{Brodsky:1983gc}.  Here $f(q)$ is the integrand 
within the diagram which includes the gluon.  In this way, $\alphastar$ absorbs 
much of the effect of these higher-order corrections, and represents 
the true strength of this gluon's coupling to the rest of the diagram.
When $f(q)$ is sensitive to large $q$,
expanding $\alphas$ on both sides at a common scale $\mu$ 
gives a form suitable for numerical calculation, 
\begin{align}
\label{firstorder}
\logqsm &= {\vev{f(q) \logqm} \over \vev{f(q)}} \\
        &\equiv \Vev{\logqm}\; , \nonumber
\end{align}
with $q^*$ given by the average log of the momentum running
through that gluon~\cite{Lepage:1993xa}.

\section{The Problem}

When $\vev{f(q)}$ vanishes, \eq{firstorder} is meaningless.
Even when nonzero, if $\vev{f(q)}$ is anomalously small compared to
$\vev{f(q) \logqm}$, the result can be misleading.  This occurs in
a variety of processes, and for some set of parameters in most
processes.  For example, it occurs for the perturbative relation between 
the bare lattice quark mass in NRQCD and its pole mass at the value 
relevant for charm.  A gluonic contribution for which $\vev{f(q)}\rightarrow 0$
is dominated by its second-order term;  \eq{firstorder}
fails because it attempts to match a first-order statement of the general
prescription to a process which is properly second-order.

\section{The Solution}

\begin{figure}[hbt]
\begin{center}
\epsfig{file=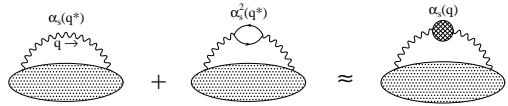, height=0.60in}
\end{center}
\caption{ 
  The BLM prescription applied to second order in $\alphastar$. }
\label{fd3}
\end{figure}

The solution is to apply the prescription beyond leading order, as illustrated
in \fig{fd3}.  This gives a reasonable $q^*$ when the leading diagram is 
anomalously small, and it approaches a sensible limit when it vanishes
altogether.  In rare cases when both the first and second order 
contributions are small, it is simple to carry this one loop higher.

The result of matching at a common scale $\mu$ is~\cite{bigpaper}
\begin{equation}
\logqsm = \Vev{\logqm} \pm \left[-\sigma^2\right]^{1/2} \; ,
\label{secondorder}
\end{equation}
with
\begin{equation}
\sigma^2 \equiv \Vev{\logqmtwo} - \Vev{\logqm}^2 \; .
\end{equation}
When $f(q)$ does not change sign, it may be interpreted as a probability
distribution, so $\sigma^2 > 0$, and \eq{firstorder} gives the correct scale.  
However, when $\sigma^2 < 0$, there are
evidently significant sign changes, and $\vev{f}$ will be anomalously
small.  In this case, \eq{secondorder} gives the correct choice of
scale.  It causes the leading-order discrepancy between the
left and right sides in \fig{fd3} to vanish, and minimizes it at next
order~\cite{bigpaper}.  Applying \eq{secondorder} requires only the 
additional computation of the average log squared within the same integrand
as in \eq{firstorder}.

\section{A Simple Example}

\begin{figure}[hbt]
\input{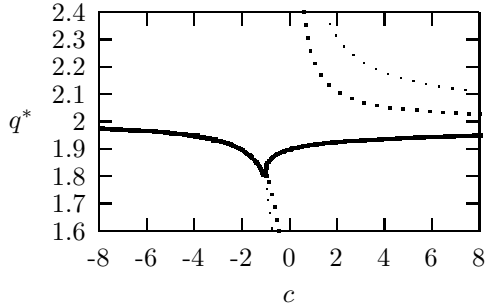}
\caption{
The BLM scale $q^*$ as a function of $c$, 
with $q_a=2.0$ and $q_b=1.8$.  The first order solution 
determines $q^*$ for $c<-1$, second order for $c>-1$.  Dark dotted lines show 
the first-order solution in regions in which it does not apply; light dotted 
lines display inapplicable second-order solutions.}
\label{deltaplot}
\end{figure}

Consider as a simple model for the integrand
\begin{equation}
 f(q) = (1+c) \delta(q - q_a) - \delta(q - q_b)\; ,
\end{equation}
sensitive to two scales, $q_a$ and $q_b$.  It suffers from
partial cancellations when $c > -1$, and $\sigma^2$ becomes
negative; when $c=0$, $\vev{f}$ vanishes
identically and \eq{firstorder} blows up.  
In \fig{deltaplot} we show the result of applying \eq{secondorder}
in this region, and \eq{firstorder} when $c < -1$ and $\sigma^2 > 0$.
It clearly behaves as expected, with $q^*$ approaching $q_a$ when $|c| \gg 1$,
$q_b$ when $c=-1$, remaining between in the interim.  Even in the
region where $c$ is large and positive and \eq{firstorder} gives
a well-behaved $q^*$, \eq{secondorder} is clearly preferable.

\section{Determining $q^*$ in $\MSbar$}

\begin{figure}[hbt]
\begin{center}
\epsfig{file=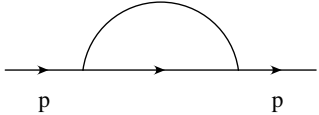, height=.27in}
\end{center}
\caption{ 
 One loop self-energy diagram in the $g \phi^3$ model.}
\label{phi3}
\end{figure}

\begin{figure}[hbt]
\input{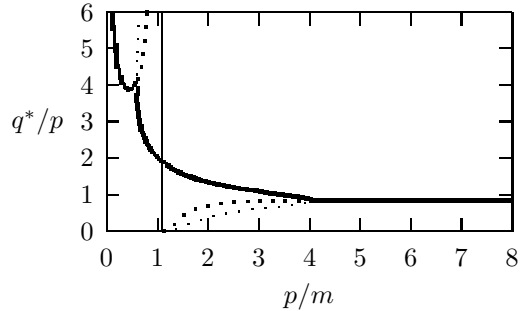}
\caption{The $\MSbar$ BLM scale $q^*/p$ as a function of momentum $p/m$ 
for the $\phi^3$ self-energy diagram. 
}
\label{plot3}
\end{figure}

The $g \phi^3$ self-energy diagram illustrated in \fig{phi3} presents a
more realistic example.  Evaluating the diagram with an additional
denominator 
\begin{equation}
 {g^2\over 2}\int {\dnk\over(2\pi)^n}\; {1\over k^2 + m^2}\;
 {1\over (p - k)^2 + m^2}\; {1\over (k^2/\mu^2)^\beta}\, ,
\end{equation} 
and expanding the result in $\beta$
produces the sequence of average logs needed to set the scale to any desired
order.  This is not much more difficult within dimensional regularization than
computing the diagram itself.  In \fig{plot3} we display the result for $q^*$
as a function of the propagator momentum $p$~\cite{bigpaper}.  The second-order 
solution is appropriate in the intermediate region.  It gives a physically reasonable 
result which connects continuously with the first-order solution appropriate
for large and small $p/m$, whereas the first-order solution diverges in this
region.

\section{Determining $q^*$ from an Existing Series}

When the terms in a series are already known, the average logs needed for
scale setting can be extracted by using the dependence on the number of
flavors $n_f$, or equivalently, $\beta_0 \equiv (11 - (2/3)n_f)/4\pi$.  
Contributions from a particular dressed gluon within a diagram
take the form 
\begin{multline}
 c_0 \alphamu + (a_1 - c_1 \bzero) \alphamutwo \\
  + (a_2 + \cdots + c_2 \bzero^2) \alphamuthree + \cdots  \; .
\end{multline}
The highest power $n$ of $\beta_0$ at each order in $\alphas$ is due to
$n$ one-loop vacuum polarization diagrams, and so at large $q$ we may
make the identification
\begin{align}
c_0 &= \vev{f} \\
c_1/c_0 &\approx \Vev{\logqm} \nonumber \\
c_2/c_0 &\approx \Vev{\logqmtwo} \; ,\nonumber 
\end{align}
allowing us to apply second-order scale setting for a series known to
$\alphas^3$.  Below we use this to present several processes for which 
\eq{secondorder} gives the appropriate scale.

\section{Higher orders}

The series coefficients from a single dressed gluon depend on
the scale $q^*$ according to~\cite{bigpaper}
\begin{multline}
\vev{f}\Big\{ \alphastar + \alphastarn{2}\big[\bzero\Delta_1\big] \\
   {}+ \alphastarn{3}\big[\bzero^2\Delta_2 + \bone\Delta_1\big] \\
   {}+ \alphastarn{4}\big[\bzero^3\Delta_3 + {5\over 2}\bzero\bone\Delta_2
            +\btwo\Delta_1\big] \\
   {}+ \alphastarn{5}\big[\bzero^4\Delta_4 + {13\over 3}\bzero^2\bone\Delta_3 \\
        {}+ 3\bzero\btwo\Delta_2 + {3\over 2}\bone^2\Delta_2 
          + \bthree\Delta_1 \big] + \cdots \Big\}\, , 
\end{multline}
with 
\begin{equation}
   \Delta_n \equiv \Vev{\left[\logqsm - \logqm\right]^n} \; .
\label{momdef}
\end{equation}
In this form, the objective in choosing an optimum scale is transparent,
regardless of the number of loops kept in $\alphas$. 
For a gluon sensitive to a narrow, large momentum region, choosing
the correct $q^*$ will approximately minimize the moments in \eq{momdef}.  
Contributions from the higher order diagrams which dress that gluon will 
be small, having been largely absorbed into the leading term.  However, 
when the region of sensitivity in momentum is large, these moments will
be minimized less effectively by a single scale $q^*$, and higher order
contributions will be larger.

The general prescription to any order is then to skip any leading terms
which are anomalously small, which will rarely be other than the first;
choose the $q^*$ which eliminates the coefficient after the first nonanomalous
term and minimizes the next; and use higher moments when available to 
check the consistency of the scale choice.

\begin{figure}[hbt]
\epsfig{file=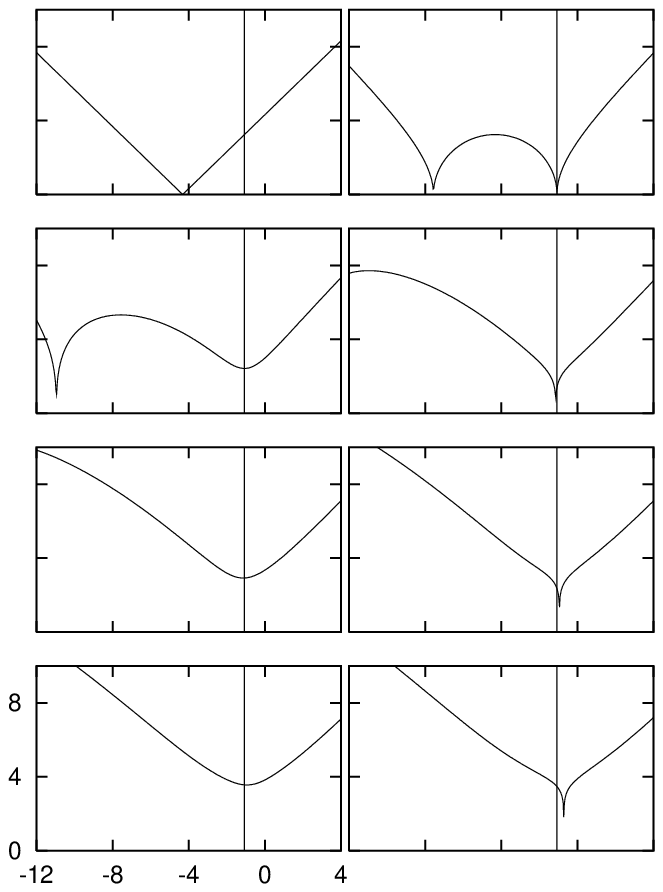, height=2.4in, width=2.5in}
\caption{The moments 
         $\left|\Delta_n\right|^{1/n}$ as functions of $\logqsm$  for 
         $n=1$ to $8$ for $\Gamma(B\rightarrow X_u e \overline{\nu}_e)$.
         The vertical line indicates the preferred scale from \eq{secondorder}.
}
\label{Bmoments}
\end{figure}

In \fig{Bmoments} we display the dependence of these moments on $\log(q^{*2})$
for the semileptonic decay width $\Gamma(B\rightarrow X_u e \overline{\nu}_e)$
expressed in terms of the $\MSbar$ $b$ mass, using results from 
\Ref{Ball:1995wa}.  In this case, \eq{secondorder} gives the appropriate 
scale, and examination of the higher moments confirms this.  Choosing the 
second-order prescription
eliminates the second moment, minimizes the third moment, and $\log(q^{*2})$
is near either the zero or minimum for all the higher moments.  As a result,
higher order coefficients are small, and the leading term effectively 
represents the strength of this gluon's coupling.

\section{Applications}

When at least the second logarithmic moment is available, we may apply
scale setting beyond leading order.  \Ref{bigpaper} presents a collection
of such series.  In \Table{aptable} we list results for four of these
for which \eq{secondorder} gives the appropriate scale:
$R_{e^+ e^-}$, the ratio of the pole to $\MSbar$ mass $M/\overline{M}$, and 
the $B$ and $t$ decay widths $\Gamma(B \rightarrow X_u e \overline{\nu})$ 
and $\Gamma(t \rightarrow b W )$ expressed in terms of $\MSbar$ masses.
These new scales represent significant corrections to the scales set
by \eq{firstorder}.
While the new scale for $M/\overline{M}$ is increased, we note 
that the range of important momenta $\Delta q$ is still relatively large, 
indicating sensitivity to
low-momentum scales even when $M$ is large, and threatening 
growing higher-order coefficients.  This is not the case for $B$ and $t$ decays
provided their series are expressed in terms of the short-distance $\MSbar$ 
masses, as indicated by their ranges $\Delta q$ and their moments, as in \fig{Bmoments}.

\begin{table*}[hbt]
\caption{Applications of second-order scale setting to several
processes.  $q_1$ gives the scale set by \eq{firstorder}; $q_2$ gives
the preferred scale by \eq{secondorder}.  
$\Delta q$, measures the range of momentum running through the gluon\protect\cite{bigpaper}.
}
\label{aptable}
\begin{tabular*}{\textwidth}{@{}l@{\extracolsep{\fill}}cccccc}

\hline 
$c_1/c_0$ & $q^*_1$ & $c_2/c_0$ & $\sigma^2$ & $q^*_2$ & $\Delta q$ \\[0.0cm]
\hline 

\multicolumn{6}{l}{
\underline{
$R_{e^+ e^-}(s)$:}}\\[.2cm]
$-.691772$ & $0.7076 \sqrt{s}$ & $-.186421$ & $-.66497$ & 
   $1.064 \sqrt{s}$ & -- \\[0.0cm]

\multicolumn{6}{l}{
\underline{
$M/\overline{M}$:}}\\[.2cm]
$-4.6862$ & $0.09603 M$ & $17.623$ & $-4.3374$ & 
   $0.27205 M$ & $0.38 M$ \\[0.0cm]

\multicolumn{6}{l}{
\underline{
$\Gamma(B \rightarrow X_u e \overline{\nu})/\Mbmsbar^5$:}}\\[.2cm]
$-4.3163$ & $0.11554 \Mb$ & $8.0992$ & $-10.531$ & 
   $0.58534 \Mb$ & $0.35 \Mb$ \\[0.0cm]

\multicolumn{6}{l}{
\underline{
$\Gamma(t \rightarrow b W )/\Mtmsbar^3$:}}\\[.2cm]
$-5.7076$ & $0.05763 \Mt$ & $6.0996$ & $-26.477$ &
   $0.75502 \Mt$ & $0.34 \Mt$ \\[0.0cm]

\hline \hline 

\end{tabular*}
\end{table*}

\begin{figure}[hbt]
\input{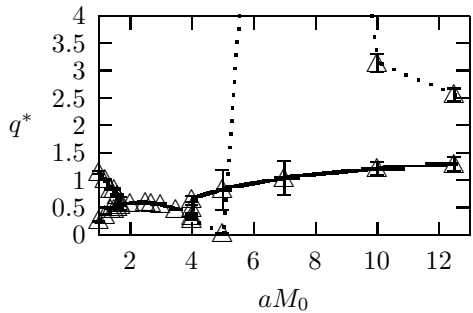}
\caption{
The BLM scale $q^*$ for the pole mass as a function of bare lattice 
mass $a M_0$. The first order solution 
determines $q^*$ between  $aM_0 = 2.00$ and $3.50$, second order elsewhere.  
Dotted lines show the first-order solution in regions in which it 
does not apply.  The point at $aM_0 = 7.00$ is off the plot.
}
\label{Mplot}
\end{figure}

In \fig{Mplot}, we show the scale associated with the series connecting the
quark pole mass to the NRQCD lattice bare mass.  At both small and large
values of the bare mass, \eq{secondorder} is appropriate and necessary to
give an accurate measure of the optimum scale. 

\section{Conclusions}

We have presented a simple general procedure for applying BLM 
scale-setting beyond leading order.  Our main result is the 
second-order prescription \eq{secondorder}, which is appropriate when 
the leading contribution is anomalously small.
We gave a general method for computing an optimum scale numerically, 
within dimensional regularization, and in cases when the coefficients 
of a series are known, and applied it to several processes. 
Finally, we discussed its application at higher orders.


\begin{thebibliography}{1}

\bibitem{Brodsky:1983gc}
S.~J.~Brodsky, G.~P.~Lepage and P.~B.~Mackenzie,
Phys.\ Rev.\  {\bf D28}, 228 (1983).

\bibitem{Lepage:1993xa}
G.~P.~Lepage and P.~B.~Mackenzie,
Phys.\ Rev.\  {\bf D48}, 2250 (1993)
[hep-lat/9209022].

\bibitem{Morningstar:1995vd}
C.~J.~Morningstar,
Nucl.\ Phys.\ B (Proc.\ Suppl.) {\bf 47}, 92 (1996)
[hep-lat/9509073].

\bibitem{bigpaper}
K.~Hornbostel {\it et al},
SMUHEP 99-12, (2000).

\bibitem{Beneke:1995qe}
M.~Beneke and V.~M.~Braun,
Phys.\ Lett.\ {\bf B348}, 513 (1995)
[hep-ph/9411229].

\bibitem{Neubert:1995vb}
M.~Neubert,
Phys.\ Rev.\ {\bf D51}, 5924 (1995)
[hep-ph/9412265].

\bibitem{Ball:1995wa}
P.~Ball, M.~Beneke and V.~M.~Braun,
Phys.\ Rev.\  {\bf D52}, 3929 (1995)
[hep-ph/9503492].

\bibitem{Ball:1995ni}
P.~Ball, M.~Beneke and V.~M.~Braun,
Nucl.\ Phys.\ {\bf B452}, 563 (1995)
[hep-ph/9502300];


\end{thebibliography}
\end{document}